\documentclass[12pt,a4paper]{article}
\DeclareFontFamily{OT1}{rsfs}{}
\DeclareFontShape{OT1}{rsfs}{m}{n}{ <-7> rsfs5 <7-10> rsfs7 <10->
rsfs10}{} \DeclareMathAlphabet{\mycal}{OT1}{rsfs}{m}{n}

\newcommand{\tl}[1]{\widetilde{#1}}
\usepackage{amsmath}

\newcommand{\tp}{\tilde p}
\newcommand{\tq}{\tilde q}
\newcommand{\tz}{\tilde z}
\newcommand{\be}{\begin{equation}}
\newcommand{\ee}{\end{equation}}
\newcommand{\beq}{\begin{eqnarray}}
\newcommand{\eeq}{\end{eqnarray}}

\usepackage{amsmath,amsthm,amsfonts}
\input amssym.def
\newsymbol\square 1003

\begin{document}

\centerline{\bf Self-dual metrics in Husain's approach}
\medskip

\centerline{M. Jakimowicz and J. Tafel}

\noindent
Institute of Theoretical Physics, University of Warsaw,
Ho\.za 69, 00-681 Warsaw, Poland, email: tafel@fuw.edu.pl

\bigskip

\noindent
{\bf Abstract}. We show that Husain's reduction of the self-dual Einstein equations  is equivalent to the Pleba\'nski equation. The B\"acklund transformation between these equations is found.  Contact symmetries of the Husain-Park equation and corresponding conservation laws are derived.

\bigskip


\null

\section{Introduction}

Probably the first application of self-dual metrics, by which we mean metrics with the self-dual Riemann tensor,  was  proposed in 1930's (see \cite{Mi} and references therein). An increased interest in this subject  started in 1970's. It was stimulated by   Pleba\'nski's reduction of the self-dual Einstein (SDE)  equations \cite{Pl},  Penrose's concept of  nonlinear graviton \cite{P} (note that a similar idea was already in \cite{Mi}) and works of Hawking and Gibbons on gravitational instantons \cite{H, GH}.

In 1988 Ashtekar et al \cite{AJS} reduced SDE equations to a system of equations for null vector fields analogous to Nahm's equations \cite{N} in the  theory of nonabelian magnetic monopoles. Chakravarty et al \cite{CMN} found explicit relations of these equations with the Pleba\'nski equation.  
Park \cite{Pa} showed that the Pleba\'nski equation follows from 2-dimensional chirial equations equiped with the Poisson bracket as a commutator. This result, in more extended form, was also obtained by Husain \cite{Hu1} in the framework of  Ashtekar's formulation. Using properties of the chiral equations he derived also an infinity of conservation laws \cite{Hu,Hu2}. 

Since the chirial equations are completely integrable, the Husain-Park equation seems specially suitable for studying completely integrable reductions of the SDE equations \cite{M,MW,DMW}. In order to create new possibilities the Husain-Park equation was recently generalized by replacing the Poisson bracket by that of  Moyal \cite{PP,PF}. 

The main purpose of this paper is to complete Husain's approach to self-duality.
By counting free functions in the Cauchy problem  Husain argued that his description  covers all self-dual metrics. This argument is not fully convincing. In section 2 we prove rigorously that, indeed, Husain's approach (with a corrected sign in the metric) is equivalent to the standard Pleba\'nski's description.  
We also find  contact symmetries of the Husain-Park equation and derive from them conserved currents which generalize those given by Husain \cite{Hu,Hu2}.

Let $M$ be a complex 4-dimensional manifold with metric $g$. In the framework of the Ashtekar approach the self-duality of the corresponding Riemann tensor can be encoded into the existence of four independent vector fields  $e_{\mu}$, $\mu=0,1,2,3$, such that \cite{AJS}
\be
[e_0,e_1]=0,\ [e_2,e_3]=0,\ [e_0,e_2]+[e_1,e_3]=0\label{1}
\ee
and 
\be
{\cal L}_{e_{\mu}}\epsilon=0,\label{2}
\ee
where ${\cal L}$ denotes the Lie derivative and $\epsilon$ is a volume form. If $\theta^{\mu}$ is the basis of 1-forms dual to $e_{\mu}$ then 
\be
\epsilon=a (\theta^0\wedge \theta^1\wedge\theta^2\wedge\theta^3),\label{3}
\ee
where $a$ is a function. Up to a constant factor the corresponding metric is defined by
\be
g=a(\theta^0\theta^2+\theta^1\theta^3).\label{4}
\ee
Thus, $e_{\mu}$ is a conformal null basis for the metric. Note that $\epsilon$ is not the invariant volume form.

Following \cite{CMN} one can relate with the vectors $e_{\mu}$ coordinates $p$, $q$, $\tp$, $\tq$ and a function $\Omega$ (denoted by $p$, $q$, $r$, $s$ and $\alpha$ in \cite{CMN}) satisfying the Pleba\'nski equation \cite{Pl}
\begin{equation}
  \Omega _{,p\tp} \Omega _{,q\tq} - \Omega _{,p\tq} \Omega _{,q \tp }=1.
  \label{5}
\end{equation}
In terms of $\Omega$ the metric tensor takes the form 
\begin{equation}
g=\Omega_{,p\tp}\,dp\,d\tp + \Omega_{,p\tq}\,dp\,d\tq + \Omega_{,q\tp}\,dq\,d\tp + \Omega_{,q\tq}\,dq\,d\tq\label{6}
\end{equation}
and the general expression for  $e_{\mu}$ is given by 
\be
e_0 =  a( u_{,q}\partial_p - u_{,p}\partial_q),\ \ e_1 = a( -v_{,q}\partial_p + v_{,p}\partial_q)  
\label{7}
\ee
\be
\begin{array}{rcl}
e_2 & = & a[(-v_{,q} \Omega _{,p\tp} + v_{,p} \Omega _{,q\tp})\partial_{\tq} +(v_{,q} \Omega _{,p\tq} - v_{,p} \Omega _{,q\tq})\partial_{\tp}]  \\
e_3 & = & a[(-u_{,q} \Omega _{,p\tp} +u_{,p} \Omega _{,q\tp})\partial_{\tq} + (u_{,q} \Omega _{,p\tq} - u_{,p} \Omega _{,q\tq})\partial_{\tp}]\ .
\label{8}
\end{array}
\ee
Here
\begin{equation}
a=(u_{,q}v_{,p}-v_{,q}u_{,p})^{-1}\label{9}
\end{equation}
and $u$, $v$ are solutions of the covariant "wave" equation 
\be 
\begin{array}{rcl}
\Omega_{,q\tq}u_{,p\tp}+\Omega_{,p\tp}u_{,q\tq}-\Omega_{,q\tp}u_{,p\tq}-\Omega_{,p\tq}u_{,q\tp}&=&0\\
\Omega_{,q\tq}v_{,p\tp}+\Omega_{,p\tp}v_{,q\tq}-\Omega_{,q\tp}v_{,p\tq}-\Omega_{,p\tq}v_{,q\tp}&=&0 
\label{10}
\end{array} 
\ee			
such that $a$ is regular. 
For $u=q$ and $v=p$ vectors (\ref{7}), (\ref{8}) simplify to
\begin{equation}
\hat e_0=\partial_p\ ,\ \  \hat e_1=\partial_q \label{10b}
\ee
\be
\hat e_2=\Omega _{,q\tl{p}}\partial_{\tl{q}} - \Omega _{,q\tl{q}}\partial_{\tl{p}}\ ,\ \ \hat e_3=- \Omega _{,p\tl{p}}\partial_{\tl{q}} + \Omega _{,p\tl{q}}\partial_{\tl{p}}. 
\label{10a}
\end{equation}     
{\bf Remark 1}.\emph{ Under assumptions (\ref{7}), (\ref{8})   equations (\ref{1}) reduce to ({10}) and  the equation
\begin{equation}
  \Omega _{,p\tp} \Omega _{,q\tq} - \Omega _{,p\tq} \Omega _{,q \tp }=h(p,q)\ ,
  \label{10c}
\end{equation}
where $h\neq 0$.
One can obtain $h=1$ by transforming $p$ and $q$.}                                                                     

In Husain's approach \cite{Hu} solutions of (\ref{1}) and (\ref{2}) have the following form in some coordinates $p'$, $q'$, $\tp'$, $\tq'$     (denoted by $t$, $x$, $p$, $q$ in \cite{H})                                                               {\setlength\arraycolsep{2pt}
\begin{equation}
e'_0 =  \partial_{p'},\ \ e'_1 = \partial_{q'} \label{11}
\ee
\begin{equation}
e'_2  = \partial_{p'} +\Lambda _{,q'\tp'}\partial_{\tq'} - \Lambda _{,q'\tq'}\partial_{\tp'}\ ,\ \ 
e'_3  =  \partial_{q'} - \Lambda _{,p'\tp'} \partial_{\tq'} + \Lambda _{,p'\tq'}\partial_{\tp'}. 
\label{12}
\end{equation}} 
Vectors $e'_{\mu}$ are independent iff
\be
a'=\Lambda_{,q'\tq'}\Lambda_{,p'\tp'}-\Lambda_{,q'\tp'}\Lambda_{,p'\tq'}\neq 0\ .\label{28}
\ee
The corresponding metric tensor is given by 
{\setlength\arraycolsep{2pt}
\begin{equation}
\begin{array}{rcl}
g&=&\Lambda_{,p'\tp'}dp'd\tp' + \Lambda_{,p'\tq'}dp'd\tq' + \Lambda_{,q'\tp'}dq'd\tp' + \Lambda_{,q'\tq'}dq'd\tq'\\&&+ 
a'^{-1}[(\Lambda_{,q'\tp'}d\tp' + \Lambda_{,q'\tq'}d\tq')^2 + (\Lambda_{,p'\tp'}d\tp' + \Lambda_{p'\tq'}d\tq')^2 ]\ .
\label{14}
\end{array}
\end{equation}
(Note that the original metric of Husain \cite{H} contains erroronous signs coming with  expressions containing  $dp'$ and $dq'$.)   

Substituting (\ref{11}), (\ref{12}) to equations (\ref{1}) yields
\begin{equation}
\Lambda_{,p'p'} + \Lambda_{,q'q'} + \Lambda_{,q'\tp'}\Lambda_{,p'\tq'} - \Lambda_{,q'
\tq'}\Lambda_{,p'\tp'}=h'(p',q')\ ,
\label{13}
\end{equation}
where $h'$ is a function of two variables. Shifting $\Lambda$ by an appropriate function of $p'$ and $q'$ transforms (\ref{13}) to the Husain equation
\begin{equation}
\Lambda_{,p'p'} + \Lambda_{,q'q'} + \Lambda_{,q'\tp'}\Lambda_{,p'\tq'} - \Lambda_{,q'
\tq'}\Lambda_{,p'\tp'}=0.
\label{13a}
\end{equation}
In terms of  potentials $A_{q'}=-\Lambda_{,p'}$, $A_{p'}=\Lambda_{,q'}$ equation 
(\ref{13a})   takes the form                      
\begin{eqnarray}
\label{15}
&&\partial_{p'}A_{q'}-\partial_{q'}A_{p'}+\left\{A_{p'},A_{q'}\right\}=0\\
\label{16}
&&\partial_{p'}A_{p'}+\partial_{q'}A_{q'}=0\ ,
\end{eqnarray} 
where \{ \} denotes  the Poisson bracket with respect to $\tp'$ and $\tq'$. 
Equations (\ref{15}), (\ref{16}) are analogous to the chirial equations (see e.g. \cite{MW}).

Since equations (\ref{5}) and (\ref{13a}) are both reductions of SDE in the form of a second order equation for one function it is very likely that they are equivalent \cite{Hu}.
Still it is possible that some descrete parameters or even functions of two variables are missing in the description of self-dual metrics via equation  (\ref{13a}). For this reason, we will investigate  explicitly a passage between the equations.

\section{Equivalence to the Pleba\'nski equation}

 Due to results of Chakravarty et al \cite{CMN} it is clear that vectors  (\ref{11}), (\ref{12}) can be represented in the form  (\ref{7}), (\ref{8}).   Let a solution $\Lambda$  of (\ref{13}), satisfying condition (\ref{28}), be given. It follows from (\ref{7}) and (\ref{11}) that  $\tp'$, $\tq'$ are independent of  $p$ and $q$ and that $p'$, $q'$ coincide with $v$ and $u$, respectively, modulo functions of $\tp$ and $\tq$. Let us assume the simplest case
\be 
p'=v,\ q'=u,\ \tp'=\tp,\ \ \tq'=\tq.\label{17}
\ee   
 Note that $p$ and $q$ are not yet defined as functions of the primed coordinates. Under (\ref{17})  vectors (\ref{11}) take the form  (\ref{7}).
Compatibility of (\ref{12}) with (\ref{8}) requires
\be
e'_2(p)=e'_3(p)=0\ ,\ \ e'_2(q)=e'_3(q)=0\label{25}
\ee
and
\be
\begin{array}{rcl}
&e'_2(\tp)=-e'_1(\Omega_{,\tq})\ ,\ &e'_3(\tp)=e'_0(\Omega_{,\tq})\\
&e'_2(\tq)=e'_1(\Omega_{,\tp})\ ,\ &e'_3(\tq)=-e'_0(\Omega_{,\tp})\ ,\label{26}
\end{array}
\ee
where $e'_{\mu}$       are given by (\ref{11}) and (\ref{12}). 
Equations (\ref{25}) define coordinates $p,\ q$. They admit nontrivial solutions since $[e'_2,e'_3]=0$. It follows from (\ref{17}) and (\ref{25}) that 
\be
\begin{array}{rcl}
\partial_{\tp}=\partial_{\tp'}-a'^{-1}(\Lambda_{,p'\tp'}\partial_{p'}+\Lambda_{,q'\tp'}\partial_{q'})\\
\partial_{\tq}=\partial_{\tq'}-a'^{-1}(\Lambda_{,p'\tq'}\partial_{p'}+\Lambda_{,q'\tq'}\partial_{q'}),\label{27}
\end{array}
\ee
where $a'$ is given by (\ref{28}).

Upon an integration and a possible change of $\Omega$, equations (\ref{26}) yield 
\be
\Lambda_{,\tp'}=\Omega_{,\tp}\ ,\ \ \Lambda_{,\tq'}=\Omega_{,\tq}.\label{20}
\ee
Equations (\ref{20}) admit solutions   with respect to $\Omega$ since the integrability condition 
\be
(\Lambda_{,\tp'})_{,\tq}=(\Lambda_{,\tq'})_{,\tp}\label{23}
\ee
is satisfied due to (\ref{27}).  

Thus, vectors (\ref{11}), (\ref{12}) can be put into the form (\ref{7}), (\ref{8}). 
Equations (\ref{10}) and (\ref{10c})  are neccessarily satisfied (see Remark 1). Hence, every solution of the Husain equation defines some solution  of the Pleba\'nski equation.
  
Consider now the reverse transformation. 
Let a solution $\Omega$ of equation (\ref{10c})  be given together with the class of  vectors (\ref{7}), (\ref{8}) depending on functions $u$, $v$ restricted by equations (\ref{10}). In order to guarantee equality of the r.h.s. of (\ref{7}) and (\ref{11}) we assume that the primed coordinates are given by (\ref{17}) with functions $u$, $v$ which are to be defined.
 Vectors (\ref{8}), when expressed in terms of the primed  coordinates, acquire components  spanned by $\partial_{q'}$, $\partial_{p'}$. They simplify to those in (\ref{12}) provided   
\be
u_{,p}=-\hat e_3(v),\ \ v_{,p}=\hat e_3(u)\label{18}
\ee
\be
u_{,q}=\hat e_2(v),\ \ v_{,q}=-\hat e_2(u),\label{19}
\ee
$\hat e_2$ and $\hat e_3$ being given by (\ref{10a}).
Conditions on  components spanned by $\partial_{\tq'}$, $\partial_{\tp'}$ in (\ref{8}) and (\ref{12}) give rise to (\ref{20}) (after an integration and an allowed transformation of $\Lambda$).

Thus, in order to construct the transformation leading from (\ref{7}) and (\ref{8}) to (\ref{11}) and (\ref{12}) one should prove the existence of solutions $u$, $v$, $\Lambda$ of equations (\ref{18}), (\ref{19}) and (\ref{20}).  
Consider (\ref{18}) as evolution equations with respect to the 'time' $p$ and (\ref{19}) as constraints. Denote $X=u_{,q}-\hat e_2(v)$ and $Y=v_{,q}+\hat e_2(u)$. A repeated use of equations (\ref{18}) and the commutation properties (\ref{1}) shows that
\be
X_{,p}=-\hat e_3(Y)\ ,\ \ Y_{,p}=\hat e_3(X).\label{21}
\ee
It follows from (\ref{21}) that constraints (\ref{19}) are preserved when $p$ changes. For a fixed initial value of $p$ equations (\ref{19}) possess solutions since they can be treated as evolution equations with respect to $q$.  Thus, solutions $u$, $v$ of equations (\ref{18}), (\ref{19}) exist and are defined by two functions of coordinates $\tq$ and $\tp$. 

Comparing the $q$-derivative of  (\ref{18}) with the $p$-derivative  of (\ref{19})  yields equations
\be
(\hat e_0\hat e_2+\hat e_1\hat e_3)u=0\ ,\ \ (\hat e_0\hat e_2+\hat e_1\hat e_3)v=0\label{22}
\ee
which coincide with the 'wave' equations ({\ref{10}). Hence, the functions $u$, $v$ can be used to define vectors (\ref{7}), (\ref{8}). It remains to show that equations (\ref{20}) admit solutions for $\Lambda$. This is true since the integrability 
condition
\be
(\Omega_{,\tp})_{,\tq'}=(\Omega_{,\tq})_{,\tp'}\label{24}
\ee
is a consequence of transformation (\ref{17}) and equations (\ref{18}), (\ref{19}). Since vectors (\ref{7}), (\ref{8})  can be put into the form (\ref{11}), (\ref{12}) the function $\Lambda$ satisfies neccessarily the Husain equation (\ref{13}). Summarizing, we obtain the following theorem.

\vspace{2mm}
\noindent
{\bf Theorem 1}.
\emph{Transformation (\ref{20}) relates solutions of the Pleba\'nski equation (\ref{10c}) with $h\neq 0$ to solutions of the Husain equation (\ref{13}) satisfying condition (\ref{28}). Equations (\ref{17}), (\ref{25})
  define unprimed coordinates in terms of primed ones and equations (\ref{17}), (\ref{18}), (\ref{19}) define the reverse transformation of coordinates.}

\vspace{2mm}

\noindent
Note that transformation (\ref{20}) is a kind of  the Backl\"und transformation (see e.g. \cite{MW}).

\section{Contact symmetries }
In this section we obtain contact symmetries of equation (\ref{13a}). Since the symmetry methods are rather standard \cite{O}, we  present final results on infinitesimal symmetries and we give more details on their integration. 

It is convenient to replace coordinates $p'$, $q'$ by $z=p'+iq'$, $\tz=p'-iq'$ and to denote $x^A=\tp',\tq'$. Then equation (\ref{13a}) and metric (\ref{13}) take the form
\be
\Lambda_{,z\tl{z}}=\frac{i}{2}\{\Lambda_{,\tz},\Lambda_{,z}\} , \label{30}
\ee
\be 
g=\Lambda_{,zA}dzdx^A+\Lambda_{,\tz A}d\tz dx^A  +(\Lambda_{,z\tz})^{-1}\Lambda_{,zA} \Lambda_{,\tz B}dx^A dx^B\ ,\label{30a}
\ee
where \{ \} is the Poisson bracket with respect to $x^A$.

Let $z,\tz,x^A,\Lambda,\Lambda_z,\Lambda_{\tz},\Lambda_A$ be coordinates of the first jet space, which is an arena space for contact transformations.  The general infinitesimal contact symmetry of (\ref{30}) is given by the following vector field
\begin{eqnarray}
v=&&h_{,\Lambda_z}{\partial_z}+\tilde h_{,\Lambda_{\tl{z}}}{\partial_{\tl{z}}}+f^A
\partial _A
+\left(\Lambda_zh_{,\Lambda_z}+\Lambda_{\tl{z}}\tilde h_{,\Lambda_{\tl{z}}}-h-\tilde h+c\Lambda+f\right){\partial _{\Lambda}}\\\nonumber
&&+(c\Lambda_{z}- h_{,z}){\partial _{\Lambda_z}}+\left(c\Lambda_{\tl{z}}-\tilde h_{,\tz}\right)\partial _{\Lambda_{\tl{z}}}+
\left(f_{,A}-f^B_{\ ,A}\Lambda_B+c\Lambda_A\right)\partial _{\Lambda_A}\label{31}
\end{eqnarray}
Here  $h=h(z,\Lambda_z)$, $\tilde h=\tilde h(\tz,\Lambda_{\tz})$ are functions of two variables and the derivatives $h_{,z}$, $h_{,\tilde z}$ are taken for fixed $\Lambda_z$ and $\Lambda_{\tilde z}$. Functions $f,f^A$ depend only on coordinates $x^B$ and
\be
f^A_{\ ,A}=c={\text const}\ .\label{32}
\ee
It is easy to see that vector field (35) generates a group of transformations such that the new coordinates $x'^A$ are given by 
\be
x'^A=F^A(x^B)  ,\label{33}
\ee
where 
\be 
det(F^A_{\ ,B})=\alpha=const\ .\label{34}
\ee
It also follows from (35) that there are functions $G,\ H,\ \tilde G,\ \tilde H$ of two variables such that
\begin{eqnarray}
z'   & = & G(z,\Lambda_z)             \ ,\ \ \Lambda'_{z'}=H(z,\Lambda_z)\ ,    
\nonumber  \\
\tz' & = & \tilde G(\tz,\Lambda_{\tz})\ ,\ \ \Lambda'_{\tz'}=\tilde H(\tz,\Lambda_{\tz})\ .
\label{35}
\end{eqnarray}
Substituting (\ref{35}) into the condition $\Lambda'_{z',\tz'}=\Lambda'_{\tz',z'}$
yields
\be
G_{,z}H_{,\Lambda_z}-H_{,z}G_{,\Lambda_z}=\tilde G_{,\tz}\tilde H_{,\Lambda_{\tz}}-\tilde 
H_{,\tz}\tilde G_{,\Lambda_{\tz}}=\beta=const\ .\label{36}
\ee
It turns out that the only remaining condition for the above transformation to be a symmetry of (\ref{30}) is 
\be
\alpha=\beta\ .\label{37}
\ee
Summarizing, we obtain the following result.

\vspace{2mm}
\noindent
{\bf Theorem 2}.
\emph{The continuous contact symmetry group of the Husain equation is given by (\ref{33})-(\ref{37}). Elements of this group preserve the metric up to the constant factor $\alpha$.}

\vspace{2mm}

\noindent

\section{Conserved currents}
It follows from (35) that variation of equation (\ref{30}) vanishes when all the coordinates are not varied and $\delta \Lambda=\epsilon F$, where $\epsilon$ is a small parameter and
\be
F=f+c\Lambda-\Lambda_{,A}f^A-h-\tilde h\ .\label{38}
\ee
This fact translates into the conservation law
\be
\partial_{\tilde z}I^{(o)}_{z}+\partial_{ z}I^{(o)}_{\tilde z}=0\ ,\label{39}
\ee
where 
\be
I^{(o)}_{z}=\partial_z F+i\{\Lambda_z,F\}\ ,\ \ I^{(o)}_{\tilde z}=\partial_{\tilde z} F-i\{\Lambda_{\tilde z},F\}\label{40}
\ee
and $\partial_z$, $\partial_{\tilde z}$ are the total derivatives. 

Let $D_z$, $D_{\tilde z}$ denote covariant derivatives such that 
\be
 I^{(o)}_{z}=D_zF\ ,\ \ I^{(o)}_{\tilde z}=D_{\tilde z}F\ .\label{41}
\ee
Following Husain \cite{Hu,Hu2} we can define a hierarchy of conserved currents $I^{(n)}$ begining with $I^{(0)}$. Let functions $F^{(n)}$, with $F^{(0)}=F$, be given by the recursive relations
\be
\partial_z F^{(n+1)}=D_z F^{(n)}\ ,\ \ \partial_{\tilde z} F^{(n+1)}=-D_{\tilde z} F^{(n)}\ .\label{42}
\ee
One can  show by induction that  functions $F^{(n)}$ exist and the currents 
\be
I^{(n)}_z=D_z F^{(n)}\ ,\ \ I^{(n)}_{\tilde z}=D_{\tilde z} F^{(n)}\label{43}
\ee 
 satisfy equation (\ref{39}).
When $F=f$ one obtains currents found by Husain \cite{Hu,Hu2}. Functions $F=c\Lambda-\lambda_Af^A$ and $F=-h-\tilde h$ yield new infinite classes of conserved currents. 

\section{Conclusions}

We have shown that any solution of the Pleba\'nski equation defines a class of solutions of the Husain equation and vice verse (Theorem 1). Vectors (\ref{11}), (\ref{12}) yield a particular solution of equations (\ref{1}), (\ref{2}), whereas expressions (\ref{7}), (\ref{8}) represent the most general solution. We have found all continuous contact transformations of the Husain-Park equation (\ref{13a}) (Theorem 2). These symmetries give rise to conserved currents (\ref{43}).

\bigskip
\noindent {\bf Acknowledgements}. We are grateful to Andrzej Trautman for indicating us ref. \cite{Mi}.
This work was partially supported by the Polish  Committee for Scientific Research (grant 1 P03B 075 29).

\end{document}